\documentclass[twocolumn,9pt]{article} % here we use the article class, rather than elsarticle

\usepackage[square,numbers,sort&compress,comma]{natbib}
\usepackage{float}
\usepackage{booktabs}
\usepackage{amsmath}
\usepackage{amssymb}
\usepackage{caption}
\usepackage{graphicx}
\usepackage{latexsym}
\usepackage{times}
\usepackage[pagewise]{lineno}
\usepackage{hyperref}
\usepackage{multirow}
\usepackage[table,xcdraw]{xcolor}

\usepackage{tikz}
\usetikzlibrary{fit,positioning,calc,shapes,arrows.meta}
\topmargin - 12pt % might need to be set to 0pt for some installations
\usepackage{xspace}   % so \Da etc. work outside $…$
\newcommand{\Da}{\ensuremath{\mathrm{Da}}\xspace}

\newcommand{\Rey}{\ensuremath{\mathrm{Re}}\xspace}
\newcommand{\Ma}{\ensuremath{\mathrm{Ma}}\xspace}
\newcommand{\Sc}{\ensuremath{\mathrm{Sc}}\xspace}

\definecolor{myBlueLink}{RGB}{70,70,200}  %%% ILLINOIS BLUE
\definecolor{myBlue}{RGB}{18,28,43}  %%% ND BLUE
\definecolor{myOrange}{RGB}{26,82,126} % monochromatic lighter
\definecolor{IllinoisOrange}{RGB}{232,74,39}  %%% ILLINOIS ORANGE
\definecolor{IllinoisBlue}{RGB}{19,41,75}   %%% ILLINOIS BLUE
\definecolor{mainText}{RGB}{18,28,43}  %%% ND BLUE
\definecolor{ForestGreen}{rgb}{0.0, 0.2, 0.13}
\definecolor{darkpastelgreen}{rgb}{0.01, 0.75, 0.24}
\definecolor{antiquewhite}{rgb}{0.98, 0.92, 0.84}

\newcommand{\bh}{\mathbf{h}}

\def \Da {\mathrm{Da}}

\def \Rey {\mathrm{Re}}
\def \Ma {\mathrm{Ma}}
\def \Pr {\mathrm{Pr}}
\def \Sc {\mathrm{Sc}}

\definecolor{lc1}{HTML}{CE282A}
\definecolor{lc2}{HTML}{255BCB}
\definecolor{lc3}{HTML}{009647} %2DCE65}
\definecolor{lc4}{HTML}{A21D8E}
\definecolor{lc5}{HTML}{F1AB65}
\definecolor{lc6}{HTML}{89B6D4}
\definecolor{lc7}{HTML}{A0A0A0}

\newenvironment{noinditemize}
{\begin{itemize}[leftmargin=*,label=\color{myOrange}$\bullet$,itemsep=0in]\raggedright}
{\end{itemize}}

\DeclareMathOperator*{\argmin}{argmin}

\renewenvironment{abstract}%
              {% - begin definition
               \small% - select font
               {\bfseries \abstractname}% - select font
               \par% - end a paragraph (skip \parsep)
               \vspace{10pt}% - add vertical space
              }% - complete definition

\renewcommand\abstractname{Abstract}

\newcommand{\nomenclature}% - name of command
              [1]% - number of arguments
              {% - begin definition
               \bgroup% - begin a local group
               \flushleft% - turn on flushleft option
               \small\bf% - select font
               #1% - insert title text
               \par% - end a paragraph (skip \parsep)
               \egroup% - terminate local group
              }% - complete definition

\renewcommand{\section}% - name of command
              [1]% - number of arguments
              {% - begin definition
               \bgroup% - begin a local group
               \flushleft% - turn on flushleft option
               \small\bf% - select font
               \refstepcounter{section}% - increment counter
               \arabic{section}. #1% - insert title text
               \par% - end a paragraph (skip \parsep)
              \egroup% - terminate local group
              }% - complete definition

\renewcommand{\subsection}% - name of command
              [1]% - number of arguments
              {% - begin definition
               \bgroup% - begin a local group
               \flushleft% - turn on flushleft option
               \small\em% - select font
               \refstepcounter{subsection}% - increment counter
               \arabic{section}.% - insert title text
               \arabic{subsection}. #1% - insert title text
               \par% - end a paragraph (skip \parsep)
               \egroup% - terminate local group
              }% - complete definition

\renewcommand{\subsubsection}% - name of command
              [1]% - number of arguments
              {% - begin definition
               \bgroup% - begin a local group
               \flushleft% - turn on flushleft option
               \small\em% - select font
               \refstepcounter{subsubsection}% - increment counter
               \arabic{section}.% - insert title text
               \arabic{subsection}.% - insert title text
               \arabic{subsubsection}. #1% - insert title text
               \par% - end a paragraph (skip \parsep)
               \egroup% - terminate local group
              }% - complete definition

  \newcommand{\acknowledgement}% - name of command
              [1]% - number of arguments
              {% - begin definition
               \bgroup% - begin a local group
               \flushleft% - turn on flushleft option
               \small\bf% - select font
               #1% - insert title text
               \par% - end a paragraph (skip \parsep)
               \egroup% - terminate local group
              }% - complete definition

  \newcommand{\sectionbib}% - name of command
              [1]% - number of arguments
              {% - begin definition
               \bgroup% - begin a local group
               \flushleft% - turn on flushleft option
               \small\bf% - select font
               #1% - insert title text
               \par% - end a paragraph (skip \parsep)
               \egroup% - terminate local group
              }% - complete definition

\begin{document}

% -------------------------------------------------------------------- %
% -------------------------------------------------------------------- %
% -------------------------------------------------------------------- %

% -------------------------------------------------------------------- %

\small
\baselineskip 10pt

% -------------------------------------------------------------------- %
% -------------------------------------------------------------------- %
% -------------------------------------------------------------------- %
\setcounter{page}{1}
% -------------------------------------------------------------------- %
\title{\LARGE \bf Solver-in-the-loop training of deep learning closures for large-eddy simulation of turbulent premixed jet flames}

\author{{\large Priyesh Kakka$^{a}$, Jonathan F. MacArt$^{a,*}$}\\[10pt]
        {\footnotesize \em $^a$Department of Aerospace and Mechanical Engineering, University of Notre Dame, Notre Dame, IN 46556, USA}\\[-5pt]
        }

\date{}  %%% Leave as is, do not add date;

% -------------------------------------------------------------------- %
% -------------------------------------------------------------------- %
% -------------------------------------------------------------------- %
\twocolumn[\begin{@twocolumnfalse}
\maketitle
\rule{\textwidth}{0.5pt}
\vspace{-5pt}

\begin{abstract} % 100 to 300 words.
Large-eddy simulation (LES) turbulence models often fail to capture the effects of chemical heat release and the resulting modulation of turbulence in premixed flames, underscoring the need for a framework that remains accurate across a broad range of physical regimes. We develop an augmented eddy-viscosity closure, based on deep neural networks calibrated jointly with the LES solution using adjoint-based optimization and differentiable programming, ensuring consistency with the governing partial differential equations (PDEs). Several objective functions and training methods are examined, and each model is assessed for its capability to interpolate and extrapolate across a wide range of Damk\"ohler numbers. Relative to the Smagorinsky-model baseline, the best neural network model improves \emph{a posteriori} errors in the LES primitive variables by 25--50\,\% and in the resolved Reynolds stress and scalar flux by more than 60\,\%. Crucially, the model generalizes across Damk\"ohler number regimes, maintaining stability and accuracy even for out-of-sample conditions. These results demonstrate that PDE-consistent deep learning closures can recover both mean fields and resolved turbulence statistics in LES of turbulent premixed flames and can therefore provide a broadly applicable framework for turbulent combustion modeling.
\end{abstract}
\vspace{10pt}

{\bf Novelty and significance statement}

\vspace{10pt}

We present the first PDE-consistent deep-learning turbulence closure for large-eddy 
simulation (LES) of turbulent premixed jet flames by embedding an untrained neural 
network directly into the governing filtered equations and training it using 
adjoint optimization and differentiable programming. This extension from prior 
RANS-based applications to LES of reacting flows introduces distinct challenges, 
including loss functions defined in terms of temporally evolving, three-dimensional, spatially filtered turbulent 
fields and intermittent loss evaluation within the corresponding adjoint 
formulation. In this LES context, we systematically evaluate different loss-function
constructions, training methodologies, and closure techniques, thereby establishing 
key modeling choices required for solver-embedded neural closures in reacting-flow 
LES. The resulting framework achieves accurate generalization across out-of-sample 
Damk\"ohler numbers and provides a foundation for extending PDE-consistent 
neural closures to complex turbulent combustion and diverse Damk\"ohler-number regimes.

\vspace{5pt}
\parbox{1.0\textwidth}{\footnotesize {\em Keywords:} Adjoint method; Turbulent premixed flames; Turbulence modeling; Large-eddy simulation; Differentiable programming}
\rule{\textwidth}{0.5pt}
*Corresponding author.
\vspace{5pt}
    \end{@twocolumnfalse}] 

% \linenumbers
\section{Introduction\label{sec:introduction}} \addvspace{5pt}

One principal difficulty in modeling reacting turbulence is that flames violate the scale-similarity and local-isotropy assumptions underlying conventional closures~\cite{frank1999measurements,veynante1997gradient,macart2018effects}. Furthermore, LES predictions exhibit marked sensitivity to filter width: as the filter scale increases, the unresolved dilatation effects intensify~\cite{macart2021damkohler}. These findings show that closures calibrated for LES of nonreacting free-shear flows cannot be directly applied to premixed flames across the Damk\"ohler number regime map. Instead, closures must dynamically adapt to turbulence–flame interactions, which motivates neural network models trained on high-fidelity direct numerical simulation (DNS) data to reconstruct the unclosed terms when conventional assumptions no longer hold.

Neural-network-based combustion models have largely focused on machine-learning-based manifold representations~\cite{perry2022co} and data-driven chemical-kinetics solvers~\cite{ji2021stiff,owoyele2022chemnode}, while studies addressing turbulent transport have primarily relied on \textit{a priori} training~\cite{yellapantula2021deep,lapeyre2019training}, which lacks dynamical consistency with \emph{a posteriori} predictions~\cite{duraisamy2021perspectives}. A more recent paradigm, a deep-learning PDE model (DPM)~\cite{sirignano2020dpm}, embeds the neural network training procedure directly into the numerical solver using adjoint-based optimization, which ensures consistency of the trained model with the governing equations. DPM has been successfully applied to LES of nonreacting turbulent flows~\cite{sirignano2020dpm,macart2021embedded,sirignano2023deep,liu2024adjoint,liu2025active}, and Kakka and MacArt~\cite{kakka2025neural} extended the framework to unsteady Reynolds-averaged Navier--Stokes (RANS) simulations of turbulent premixed flames, demonstrating the ability of learned models to generalize across a wide range of turbulent Damk\"ohler numbers.

Building on these foundations, we develop DPM-based LES closures that capture the Damk\"ohler number dependence of turbulence--flame interactions while maintaining dynamical consistency with the governing equations. Reference data are obtained from DNS of turbulent premixed jet flames spanning the nonreacting to strongly burning (``thin-flame'') regimes. The LES baseline employs a compressible solver with Smagorinsky-type closures for momentum, energy, and species transport. Single-step global chemistry  isolates turbulent transport effects, though the formulation can be extended to detailed kinetics. Model performance is evaluated using \emph{a posteriori} LES for in- and out-of-sample Damk\"ohler numbers, demonstrating that LES-embedded neural closures accurately recover both primitive fields and subfilter transport with improved generalization.

\section{Governing equations\label{sec:gov_eqns_les}}
\begin{figure}[t]
    \centering
    \includegraphics[width=0.4\textwidth]{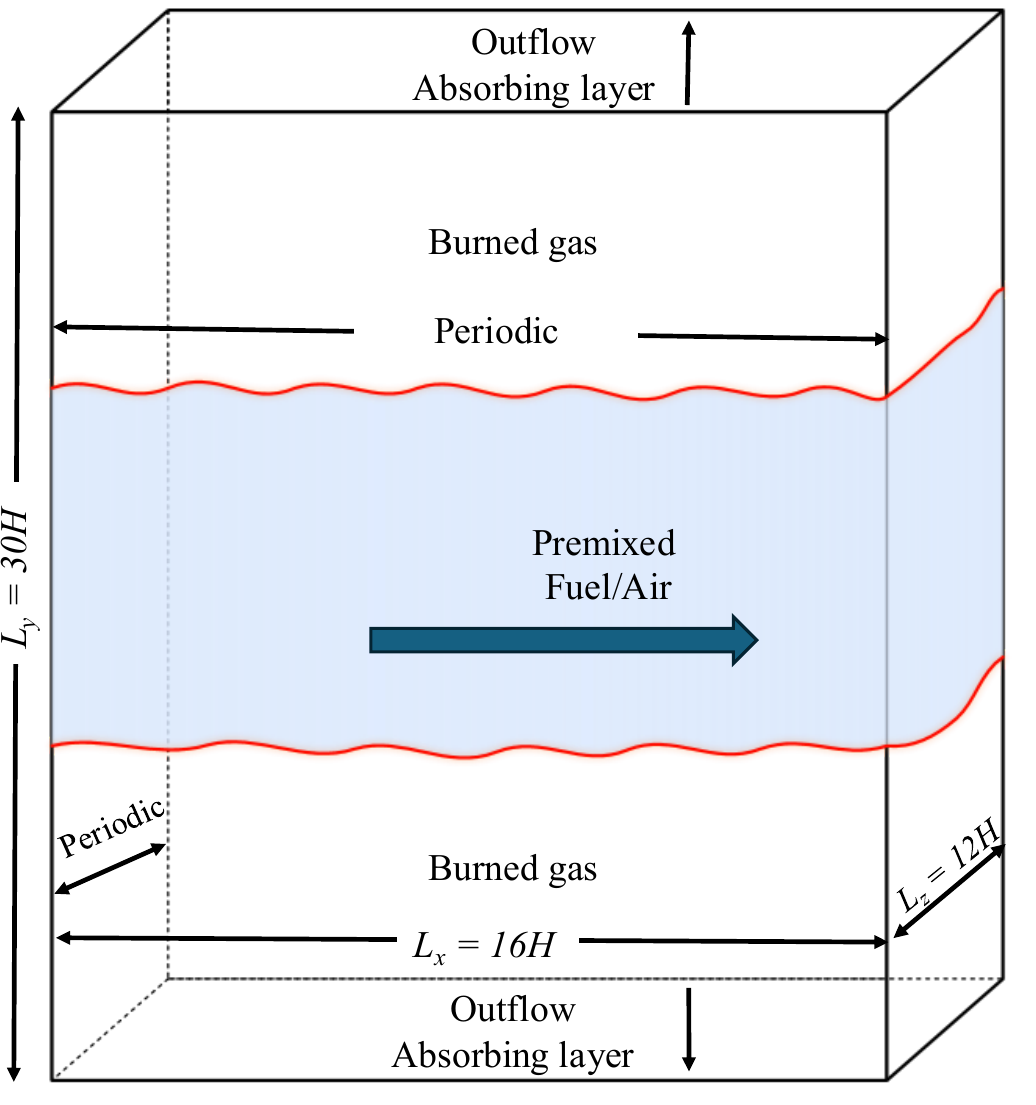}
    \caption{Schematic of the temporally evolving turbulent premixed planar jet flame DNS configuration, where $H$ is the initial slot-jet height.}
    \label{fig:dns_schematic}
\end{figure}

\addvspace{10pt}
The DNS data used for LES training follow the compressible, temporally evolving, turbulent premixed jet flames of Kakka and MacArt~\cite{kakka2025neural}, whose configuration consists of a premixed fuel/air jet core surrounded by burned gases, with flame fronts forming along the two shear layers, as shown schematically in Fig.~\ref{fig:dns_schematic}. The streamwise and spanwise directions are periodic and statistically homogeneous, and absorbing layers are applied at the cross-stream outflow boundaries. The only modification relative to the setup of Kakka and MacArt~\cite{kakka2025neural} is the initialization of localized perturbations to promote transition to turbulence, introduced as random velocity fluctuations following MacArt \textit{et al}.~\cite{macart2018effects}. The dimensionless governing equations~\cite{kakka2025neural} are simulated for scaling Reynolds and Mach numbers $\Rey=6{,}000$ and $\Ma=0.08$. Three scaling Damk\"ohler numbers are considered: $\Da_s \in \{0,\,20{,}000,\,35{,}000\}$, corresponding to turbulent Damk\"ohler numbers in the nonreacting, distributed-burning, and thin-flames regimes~\cite{kakka2025neural}. 
\subsection{Filtered LES Equations\label{subsec:filtered_eqns}}\addvspace{10pt}

LES separates the resolved and subgrid scales through the application of a spatial low-pass filter $\overline{\phi}$ with characteristic width~$\Delta$.  Favre filtering, i.e., $\widetilde{\phi} = \overline{\rho\phi}/\overline{\rho}$,  ensures proper treatment of density variations in the filtered governing equations.
Filtered fields are obtained directly from DNS data, and statistical quantities from DNS, filtered DNS (fDNS), and LES are computed by Reynolds averaging in the statistically homogeneous streamwise and spanwise directions, denoted $\langle \phi \rangle$. The dimensionless governing equations for a single-step reaction, following Kakka and MacArt~\cite{kakka2025neural}, yield the LES system expressed in terms of the conserved variables $Q = (\overline{\rho},\overline{\rho}\widetilde{u}_i,\overline{\rho}\widetilde{E},\overline{\rho}\widetilde{Y}_P)$, where $\overline{\rho}$ is the mass density, $\widetilde{u}_i$ ($i=1,2,3$) are the velocity components in the streamwise ($u$), flame-normal ($v$), and spanwise ($w$) directions,  $\widetilde{E} = \widetilde{e} + \widetilde{u}_i\widetilde{u}_i/2$ is the total energy \cite{vreman1995subgrid}, $\widetilde{e}$ is the internal energy, and $\widetilde{Y}_P$ is the product mass fraction.

\indent
Filtering the nonlinear flux terms in the governing equations introduces unclosed subgrid-scale (SGS) contributions
\begin{align}
\tau_{ij}^{\mathrm{SGS}} &= \overline{\rho}\left(\widetilde{u_i u_j} - \widetilde{u}_i\widetilde{u}_j\right),
\label{eq:sgs_stress}\\
F^{\mathrm{SGS}}_{T,j} &= \overline{\rho}\left(\widetilde{u_j T} - \widetilde{u}_j\widetilde{T}\right),
\label{eq:sgs_heat}\\
F^{\mathrm{SGS}}_{Y_{P,j}} &= \overline{\rho}\left(\widetilde{u_j Y_P} - \widetilde{u}_j\widetilde{Y}_P\right),
\label{eq:sgs_scalar}
\end{align}
representing the SGS stress tensor, SGS heat flux, and SGS scalar flux.
The unclosed SGS terms are modeled using an eddy-viscosity assumption, where the deviatoric part of the SGS  stress tensor in~\eqref{eq:sgs_stress} is closed using the Boussinesq hypothesis~\cite{pope2000turbulent},
\begin{equation}
\label{eq:subgrid_smag_closure}
\tau^{\mathrm{SGS,dev}}_{ij}
= -2\,\mu_t(Q)
\left( \widetilde{S}_{ij} - \frac{1}{3}\,\delta_{ij}\,\widetilde{S}_{kk} \right),
\end{equation}
where $\widetilde{S}_{ij} = \tfrac{1}{2}\left(\tfrac{\partial \widetilde{u}_i}{\partial x_j} + \tfrac{\partial \widetilde{u}_j}{\partial x_i}\right)$ is the resolved strain-rate tensor, and
the SGS heat and scalar fluxes are closed using the gradient-diffusion hypothesis,
\begin{align}
\label{eq:temp_closure_eq}
F^{\mathrm{SGS}}_{T,j} &= -\frac{\mu_t(Q)}{\Pr_t}\,\frac{\partial \widetilde{T}}{\partial x_j}, \\[3pt]
\label{eq:species_closure_eq}
F^{\mathrm{SGS}}_{Y_{P,j}} &= -\frac{\mu_t(Q)}{\Sc_t}\,\frac{\partial \widetilde{Y}_P}{\partial x_j},
\end{align}
where $\Pr_t=0.9$ is the constant turbulent Prandtl number~\cite{baurle2017hybrid}, and $\Sc_t=0.65$ is the constant turbulent Schmidt number~\cite{macart2018effects}.  

The baseline eddy viscosity $\mu_t$ used for comparison is obtained using the classical Smagorinsky closure~\cite{smagorinsky1963general}, which relates $\mu_t$ to the local strain-rate magnitude $|\widetilde{S}| = \sqrt{2\,\widetilde{S}_{ij}\widetilde{S}_{ij}}$ through a model coefficient $C_s$,
\begin{equation}
\label{eg:mut_smag}
\mu_t^{\textsc{sm}}(Q) = \overline{\rho}\,(C_s \Delta)^2\,|\widetilde{S}|.
\end{equation}
We use $C_s=0.3$ for stable predictions across our range of scaling Damk\"ohler numbers.  

Grid-filtered LES is performed on grids downsampled $16\times$ from the DNS mesh  ($\Delta_{16}$), resulting in LES grid size $(N_{x,\mathrm{LES}},\,N_{y,\mathrm{LES}},\,N_{z,\mathrm{LES}})=(65,\,51,\,49)$.  
A curvilinear coordinate system is used \cite{hickling2024large,liu2025active}, in which the stretched physical domain $(x,y,z)$ is mapped smoothly onto a uniform computational space; further details are available in \cite{kakka2025neural}.
Time integration is performed with a constant timestep $
  \Delta t_{\mathrm{LES}} = 8\,\Delta t_{\mathrm{DNS}} = 3.0 \times 10^{-3},
$
corresponding to a Courant–Friedrichs–Lewy (CFL) number of 0.20.  
The LES advances from $t=14.985$ to $t=16.425$ in ($N_T = 480$) timesteps with flow fields saved every 20 steps for use in adjoint-based optimization, described subsequently.

\section{Neural Network-based Closures
\label{sec:NN_closure_les}}\addvspace{10pt}

To improve modeling fidelity, an eddy-viscosity model denoted ``ME'' is applied \cite{kakka2025neural}, which for the present application augments the baseline Smagorinsky closure by embedding a neural network $\mathcal{N}$ within the LES solver to close the filtered governing equations.  
The network predicts an augmented eddy viscosity $\mu_t(Q;\theta)$, appearing in~\eqref{eq:subgrid_smag_closure} and~\eqref{eq:species_closure_eq}, and a distinct $\mu_{t,E}(Q;\theta)$ for the subgrid-scale heat flux~\eqref{eq:temp_closure_eq}. 
It is evaluated pointwise on the computational mesh using Galilean-invariant gradient features as inputs:
\begin{align}
\vspace{-2em}
  \mu_t(Q,\theta),\;&\mu_{t,E}(Q,\theta) 
  \;\;\leftarrow\;\;
  \mathcal{N}\!\Bigg(
  \frac{\partial \overline{\rho}}{\partial x_j},\,
  \frac{\partial \widetilde{u}_i}{\partial x_j},\,
  \frac{\partial \widetilde{e}}{\partial x_j},\, \nonumber\\
  &\quad
  \frac{\partial \widetilde{Y}_P}{\partial x_j},\,
  \frac{\partial \overline{p}}{\partial x_j},\,
  \mu_t^{\textsc{sm}},\,
  \Da_s\,\overline{\dot{\omega}}_P,\,
  t
  \Bigg),
  \label{eq:ME_NN}
\end{align}
where the inputs also include the baseline Smagorinsky eddy viscosity and the local chemical source term $\Da_s\,\dot{\omega}_P$, evaluated from the resolved LES fields and included primarily as a regime indicator to help the network distinguish among the different Damk\"ohler-number cases.
In practice, $\mathcal{N}$ is evaluated using local features at each cell and its six nearest neighbors, with $\mu_t^{\textsc{sm}}$ obtained from~\eqref{eg:mut_smag}. The neighboring-cell inputs provide compact stencil-level context, allowing the pointwise network evaluation to account for local spatial variation. The network is implemented as a multilayer perceptron (MLP) with four fully connected layers and 100 hidden units each, yielding 35,102 trainable parameters.  
All layers employ Gaussian Error Linear Unit (GELU) activation functions~\cite{hendrycks2016gaussian}. To avoid destabilizing the forward solves at random initialization, the untrained neural-network outputs are empirically scaled before being added to the baseline closure. In the ME formulation, the two output channels are multiplied by a tunable factor of $10^{-5}$. This localized ME formulation ensures compatibility with the LES solver while remaining lightweight for deployment in large-scale simulations.

In addition to the augmented eddy-viscosity model, we test a direct closure (DC) of the subgrid transport terms. This model directly predicts the SGS stress tensor $\tau^{\mathrm{SGS}}_{ij}(Q;\theta)$, turbulent heat flux $F^{\mathrm{SGS}}_{T,j}(Q;\theta)$, and scalar flux $F^{\mathrm{SGS}}_{Y_{P,j}}(Q;\theta)$, using the same inputs to $\mathcal{N}$ provided by \eqref{eq:ME_NN}. The predicted tensor is then projected onto its deviatoric component,
\begin{equation}
  \tau^{\mathrm{SGS,dev}}_{ij}(\theta) = \tau^{\mathrm{SGS}}_{ij}(\theta) - \tfrac{1}{3}\delta_{ij}\,\tau^{\mathrm{SGS}}_{kk}(\theta).
  \label{eq:sgs_deviatoric}
\end{equation}
Compared to the ME closure, the DC closure outputs fifteen terms rather than two, increasing the model dimensionality to 37,415 parameters with marginally higher training cost. This additional flexibility permits the network to represent anisotropy and counter-gradient fluxes but also increases the challenge of solving the optimization problem. In contrast to the ME formulation, the DC formulation remain stable at initialization without additional output scaling.

\section{Adjoint-based Optimization\label{sec:DPM-LES}}\addvspace{10pt}
The neural-network closure introduced in Section~\ref{sec:NN_closure_les} is trained 
within the LES solver by solving a PDE-constrained optimization problem that 
enforces the governing filtered equations while assimilating filtered DNS 
data at discrete time intervals. The overall adjoint-based training procedure 
is similar to that of Kakka and MacArt~\cite{kakka2025neural}, in which the 
closure model is coupled directly to the RANS solver and optimized using 
solver-consistent adjoint gradients. In this work, the framework is adapted to the 
LES formulation by (i) eliminating the need to backpropagate over auxiliary transport equations, 
(ii) evaluating the loss function $J$ only at the end of each optimization 
window, rather than at every time step, and (iii) modifying the adjoint 
formulation, resulting in a different terminal initialization of the adjoint 
variables. With these adaptations, let $\bh$ denote the residuals of the 
governing equations.
\begin{equation}
\bh(Q,\dot Q;\theta) = \bigl[h_{\overline{\rho}},\;h_{\overline{\rho}\widetilde{u}},\;h_{\overline{\rho}\widetilde{v}},\;h_{\overline{\rho}\widetilde{w}},\;h_{\overline{\rho}\widetilde{E}},\;h_{\overline{\rho}\widetilde{Y}_P}\bigr]^\top,
\end{equation}
where $\dot{Q}$ denotes the time derivative of the LES vector of conserved variables. We solve a constrained minimization problem,
\begin{equation}
  \argmin_{\theta}\; \bar J\bigl(Z(Q(\theta))\bigr)
\quad \text{subject to} \quad \bh(Q,\dot Q;\theta)=0,
\end{equation}
where $\bar J$ is a cumulative objective function, and $Z$ represents quantities of interest derived from $Q$. The discrepancy evaluated at the end of optimization window $k$ between LES predictions $Z^k$ and the corresponding fDNS targets $Z^{e,k}$ is given by
\begin{equation}
J^k = \sum_r \left(S_{r}\left(Z_r^k - Z_r^{e,k}\right)\right)^2,
\label{eq:J}
\end{equation}
where $S_{r}$ are weighting factors, described subsequently, which we use to ensure that the various terms contribute approximately equally to the objective function. 
The cumulative objective over an epoch of length $N_T$ steps is defined as the sum of these discrepancies over all optimization windows,
\begin{equation}
\bar{J}(Z(Q(\theta))) = \sum_{k=1}^{N_T/N_{\mathrm{op}}} J^k.
\label{eq:time_integrated_loss_les}
\end{equation}
where each epoch advances the LES for $N_T$ total time steps partitioned into windows of length $\tau_{\mathrm{op}}=N_{\mathrm{op}}\Delta t$, which corresponds to the time horizon between fDNS snapshots.  
Different choices of $Z(Q)$ are introduced and examined in subsequent sections to assess their influence on model convergence and predictive accuracy. 

To compute the parameter gradients for a given optimization window, we introduce the Lagrangian
\begin{equation}
\mathcal{L}^k = J^k(Z(Q)) + \check\chi^\top \bh(Q,\dot Q;\theta),
\label{eq:lagrangian}
\end{equation}
where 
\begin{equation}
\check\chi = \bigl[\check\Omega_{\overline{\rho}},\;\check\Omega_{\overline{\rho}\widetilde{u}},\;\check\Omega_{\overline{\rho}\widetilde{v}},\;\check\Omega_{\overline{\rho}\widetilde{w}},\;\check\Omega_{\overline{\rho}\widetilde{E}},\;\check\Omega_{\overline{\rho}\widetilde{Y}_P}\bigr]^\top
\end{equation}
are the  Lagrange multipliers of the forward system.
The Lagrangian reduces to $J^k$ when the PDE constraint is satisfied. Differentiating \eqref{eq:lagrangian} with respect 
to $\theta$, applying the chain rule, and requiring terms multiplying $\partial Q/\partial\theta$ to vanish yields the adjoint equation
\begin{equation}
\label{eq:chi_hat_ode}
  \frac{d\check\chi^\top}{dt} = \check\chi^\top \frac{\partial \bh}{\partial Q},
\end{equation}
integrated backward in time over an optimization window 
$t_w \in [(k-1)\tau_\mathrm{op},\,k\tau_\mathrm{op}]$ 
with $k=1,\dots,N_T/N_\mathrm{op}$. The adjoint is initialized with a terminal condition
\begin{equation}
\check\chi^{k\tau_\mathrm{op}} =
\left.\frac{\partial J^k}{\partial Q}\right|_{t=k\tau_\mathrm{op}},
\end{equation}
where $\partial J^k/\partial Q$ is obtained analytically from the objective function, and 
$\partial \bh/\partial Q$ (and $\partial \bh/\partial \theta$, used subsequently) are assembled by 
algorithmic differentiation over the flow solver \cite{hickling2024large,liu2025active}.
Consistency in residual evaluation is ensured using solution checkpointing, in which the LES state trajectory over window $k$, denoted $\lambda^k$, is stored during the forward run and retrieved during the backward pass to evaluate $\bh(Q,\theta)$ and its derivatives along the primal trajectory. The parameter gradient is then computed as 
\begin{equation}
\label{eq:grad_L} 
\nabla_\theta J^k
=
\sum_{l=1}^{N_\mathrm{op}}
\check\chi_{l}^{\!\top}
\left.\frac{\partial \bh}{\partial \theta}\right|_{l}\Delta t.
\end{equation}
Each epoch thus consists of $N_T/N_\mathrm{op}$ sequential forward (LES) and backward (adjoint) integrations corresponding to optimization windows. In each window, checkpointed states $\lambda^k$ are used to compute gradient contributions in the backward pass, and the optimized LES state initializes the subsequent window.
\subsection{Objective Functions\label{subsec:objective_functions}}\addvspace{10pt}

We now evaluate several objective functions to assess model performance across varying levels of statistical fidelity. The first variant targets the primitive variables $Z_{P} = (\bar{\rho},\,\widetilde{u},\,\widetilde{v},\,\widetilde{w},\,\widetilde{T},\,\widetilde{Y}_P)$
and enforces fidelity in the LES-resolved quantities,
\begin{equation}
J_{P}^k = \sum_{r=1}^{6} \left(S_{P,r}\left(Z_{P,r}^k - Z_{P,r}^{e,k}\right)\right)^2,
\label{eq:J_prim}
\end{equation}
where $S_P = (1,\,1,\,1,\,1,\,0.083,\,1)$ balances the relative weight of the primitive variables.

To improve physical consistency, we define an augmented objective function using second-order statistics. The resolved Reynolds stresses are defined as  
\begin{equation}
R_{ij}(y,t) =
\frac{\langle \overline{\rho}\widetilde{u}_i\widetilde{u}_j\rangle}{\langle \overline{\rho}\rangle}
- \frac{\langle \overline{\rho}\widetilde{u}_i\rangle\langle \overline{\rho}\widetilde{u}_j\rangle}{\langle \overline{\rho}\rangle^2},
\label{eq:Rij}
\end{equation}
where $i,j \in \{1,2,3\}$. For the present temporally evolving jet, the Reynolds average $\langle \cdot \rangle(y,t)$ is taken over the statistically homogeneous streamwise and spanwise directions, giving statistics that are functions of the flame-normal coordinate $y$ and time. The corresponding stress-based objective is represented using a stacked component vector $Z_R^k = (d_{uu}^k,\,d_{uv}^k,\,d_{uw}^k,\,d_{vv}^k,\,d_{vw}^k,\,d_{ww}^k)$, whose individual entries quantify the normalized discrepancy between LES and fDNS resolved Reynolds stress components according to  
\begin{equation}
d_{ij}^k \;=\;
\frac{\sum_{m=1}^{N_{y,\mathrm{LES}}}\!\big(R_{ij}^{\mathrm{LES}}(y_m,t)-R_{ij}^{\mathrm{fDNS}}(y_m,t)\big)^2}
     {\sum_{m=1}^{N_{y,\mathrm{LES}}}\!\big(R_{ij}^{\mathrm{fDNS}}(y_m,t)\big)^2}.
\label{eq:JR_loss}
\end{equation}
Using these component errors, the resolved Reynolds stress loss mirrors the structure of the primitive-field objective and is written as  
\begin{equation}
J_{\mathrm{stress}}^k
=
\sum_{r=1}^{6}\!S_{R,r}\left(Z_{R,r}^k\right),
\label{eq:JR_stress_vector}
\end{equation}
with $S_R=(10^{4},\,10^{4},\,10^{3},\,10^{4},\,5\times10^{3},\,5\times10^{3})$.  
A reduced variant isolating the flame-normal component is given by  
\begin{equation}
J_{R_{vv}}^k \;=\; 10^{4}\,\big(d_{vv}^k\big),
\label{eq:JR_vv_vector}
\end{equation}
which corresponds to selecting the $(vv)$ entry of $Z_R^k$ and directly penalizes discrepancies in the flame-normal resolved stress component.
Analogously, the resolved Favre scalar fluxes are defined as  
\begin{equation}
F_{Y_{P,j}}(y,t) \;=\; 
\frac{\langle \overline{\rho}\,\widetilde{u}_j \widetilde{Y}_P \rangle}{\langle \overline{\rho} \rangle}
- \frac{\langle \overline{\rho}\,\widetilde{u}_j \rangle \langle \overline{\rho}\,\widetilde{Y}_P \rangle}{\langle \overline{\rho} \rangle^2},
\end{equation}
where $j \in \{1,2,3\}$. The scalar-flux objective employs $Z_F^k = (d_u^k,\,d_v^k,\,d_w^k)$, where each component $d_j^k$ represents the normalized mean-squared flux difference,  
\begin{equation}
  d_{j}^k \;=\;
  \frac{\displaystyle \sum_{m=1}^{N_{y,\mathrm{LES}}}\!
  \big(F_{Y_{P,j}}^{\mathrm{LES}}(y_m,t)-F_{Y_{P,j}}^{\mathrm{fDNS}}(y_m,t)\big)^2}
       {\displaystyle \sum_{m=1}^{N_{y,\mathrm{LES}}}\!
  \big(F_{Y_{P,j}}^{\mathrm{fDNS}}(y_m,t)\big)^2},
  \label{eq:J_flux}
\end{equation}
and the total scalar-flux loss is formulated in an analogous form as  
\begin{equation}
J_{\mathrm{flux}}^k
=
\sum_{r=1}^{3}\!S_{F,r}\,\left(Z_{F,r}^k\right),
\label{eq:JF_flux_loss}
\end{equation}
where the scaling vector $S_F=(10^{2},\,10^{2},\,10^{2})$ assigns equal weighting to the three flux components. The coefficients $S_{P,r}$ in~\eqref{eq:J_prim}, $S_{R,r}$ in~\eqref{eq:JR_stress_vector}, and $S_{F,r}$ in~\eqref{eq:JF_flux_loss} are weighting factors that assign relative importance to the primitive-variable, resolved-stress, and scalar-flux objective components, respectively. In the present loss construction, the relative contribution of the second-order-statistics loss is smaller, so insufficient weighting of this term would bias the optimization toward primitive-variable accuracy and reduce the model's ability to predict the resolved stresses. Thus, the weights prevent any single component from dominating the composite objective and were tuned empirically. The combined second-order contribution to the training objective is expressed as  
\begin{equation}
    J^k_{\mathrm{SF}} = J^k_{\mathrm{stress}} + J^k_{\mathrm{flux}}.
\end{equation}

For model training, three objective variants are formulated to progressively incorporate higher-order information within the optimization framework. The first variant, ``P,'' includes only the primitive-field loss $J_P^k$, enforcing fidelity in the resolved mean quantities. The second variant, ``PRvv,'' augments this baseline with the flame-normal Reynolds stress term $J_{R_{vv}}^k$, adding sensitivity to cross-stream momentum transport. The third variant, ``PSF,'' further extends the formulation by incorporating the complete second-order term $J_{\mathrm{SF}}^k$, thereby enforcing simultaneous accuracy in both momentum and scalar transport statistics. These three objective variants, summarized in Tab.~\ref{tab:loss_var}, are trained under identical solver and optimization configurations to ensure that performance differences arise solely from the choice of the objective.
The resulting trained models are denoted ``$\mathrm{ME}x$–variant,'' where $x=\Da_s/10{,}000$ prescribes the training Damk\"ohler number, and ``variant'' indicates the training objective function.

\begin{table}[h]
\centering
\caption{Objective function variants used for training.}
\label{tab:loss_var}
\begin{tabular}{ll}
\toprule
\textbf{Variant} & \textbf{Definition} \\
\midrule
P       & $J_{P}^k$ \\
PRvv    & $J_{PR_{vv}}^k = J_{P}^k + J_{R_{vv}}^k$ \\
PSF     & $J_{PSF}^k = J_{P}^k + J_{\mathrm{SF}}^k$ \\
\bottomrule
\end{tabular}
\end{table}

\section{Training Methodology\label{sec:training}}\addvspace{10pt}

The neural closure models introduced in Section~\ref{sec:NN_closure_les} are trained within the LES framework using adjoint-based gradients derived from the optimization procedure described in Section \ref{sec:DPM-LES}. Three distinct training strategies are formulated to assess model robustness with respect to temporal sampling, data availability, and cross-regime generalization: series window training, parallel window training, and parallel data training, described subsequently. In all cases, the \emph{Adam} optimizer \cite{kingma2014adam} is used with an initial learning rate of $\alpha=10^{-3}$; this is reduced by a factor of two whenever the objective \eqref{eq:time_integrated_loss_les} fails to decrease over four consecutive epochs, and training is terminated once the learning rate falls below $10^{-7}$ or after 400 epochs. All hyperparameters are selected manually through a limited set of preliminary training runs, with emphasis on stable \emph{a posteriori} LES integration and validation accuracy.

The optimization window length for training ($N_\mathrm{op}$) is a crucial hyperparameter for successful optimization using adjoint-based methods \cite{liu2024adjoint}. For the present deep learning based closures, the choice of $N_\mathrm{op}$ must be sufficiently large to observe several time scales of the grid-scale eddies but not so large that the Lyapunov divergence of the adjoint variables (due to the chaotic nature of the turbulent flow) contaminates the computed gradients. For the present flow regimes (Reynolds and Damk\"ohler numbers), LES grid-filter size, and objective functions, we found  $N_{\mathrm{op}}=60$ to be a stable compromise across all training strategies and cases.
\vspace{1 cm}
\subsection{Series Window Training\label{sec:series_training}}\addvspace{10pt}

In this baseline configuration, the neural network is updated after every optimization window of length $N_{\mathrm{op}}$ steps. Thus, each epoch consists of $N_T/N_{\mathrm{op}}$ sequential parameter updates, where the forward pass provides checkpointed states for the adjoint calculation and the backward pass yields the gradient contribution of each window. The optimized LES state at the end of each window is then used to restart the simulation for the subsequent optimization window.

\subsection{Parallel Window Training\label{sec:parallel_window_training}}\addvspace{10pt}

This strategy modifies the optimization by re-initializing the LES after each optimization window with filtered DNS fields rather than advancing continuously from one window to the next  \cite{sirignano2020dpm,macart2021embedded,sirignano2023deep,hickling2024large,liu2025active}. Thus, each window begins from an fDNS (i.e., target) state and evolves forward over  a time window of size $\tau_{\mathrm{op}} = N_{\mathrm{op}}\Delta t$, after which the adjoint is integrated backward over the same horizon to accumulate parameter gradients.
We use four GPUs to evolve four distinct training windows in parallel, with one optimization window assigned to each GPU. The resulting gradients are averaged across devices to synchronously update the shared network parameters. Models trained with this approach are denoted ``ME$x$-variant-PW'' (parallel window).

\subsection{Parallel Data Training\label{sec:parallel_model_training}}\addvspace{10pt}

Finally, to evaluate scalability with respect to multiple operating regimes, a third strategy optimizes simultaneously for two scaling Damk\"ohler numbers, $\Da_s \in \{0,\,35{,}000\}$ using a shared neural network and series window training to capture temporal flow evolution. Separate LES predictions are made in parallel for each $\Da_s$, and the gradients obtained from the two cases are averaged across different GPUs. The shared parameters are then updated using this aggregated sensitivity, with the learning rate adaptively set to the smaller of the two values to ensure stable convergence. Models trained in this fashion are denoted as ``ME(0,35)-variant.''

\section{Results \label{sec:results}}\addvspace{10pt}

The performance of the neural closures is assessed across different training 
strategies, objective formulations, and scaling Damk\"ohler numbers. The loss 
is evaluated every $20$ LES time steps, and the cumulative objective 
$\bar J$ (see Section~\ref{subsec:objective_functions}) provides the basis for all 
percentage improvements reported. Results are presented in three parts: 
Section~\ref{subsec:Closure Model and Training Method Evaluation} evaluates 
closure formulations and training methods; 
Section~\ref{subsec:Effect of Loss Formulation} compares alternative loss 
functions; 
Section~\ref{subsec:In-Sample and Out-of-sample Predictions with ME20-PRvv} 
examines performance of models trained at different scaling Damk\"ohler numbers.
\begin{figure}[b]
    \centering
\includegraphics[width=0.95\linewidth]{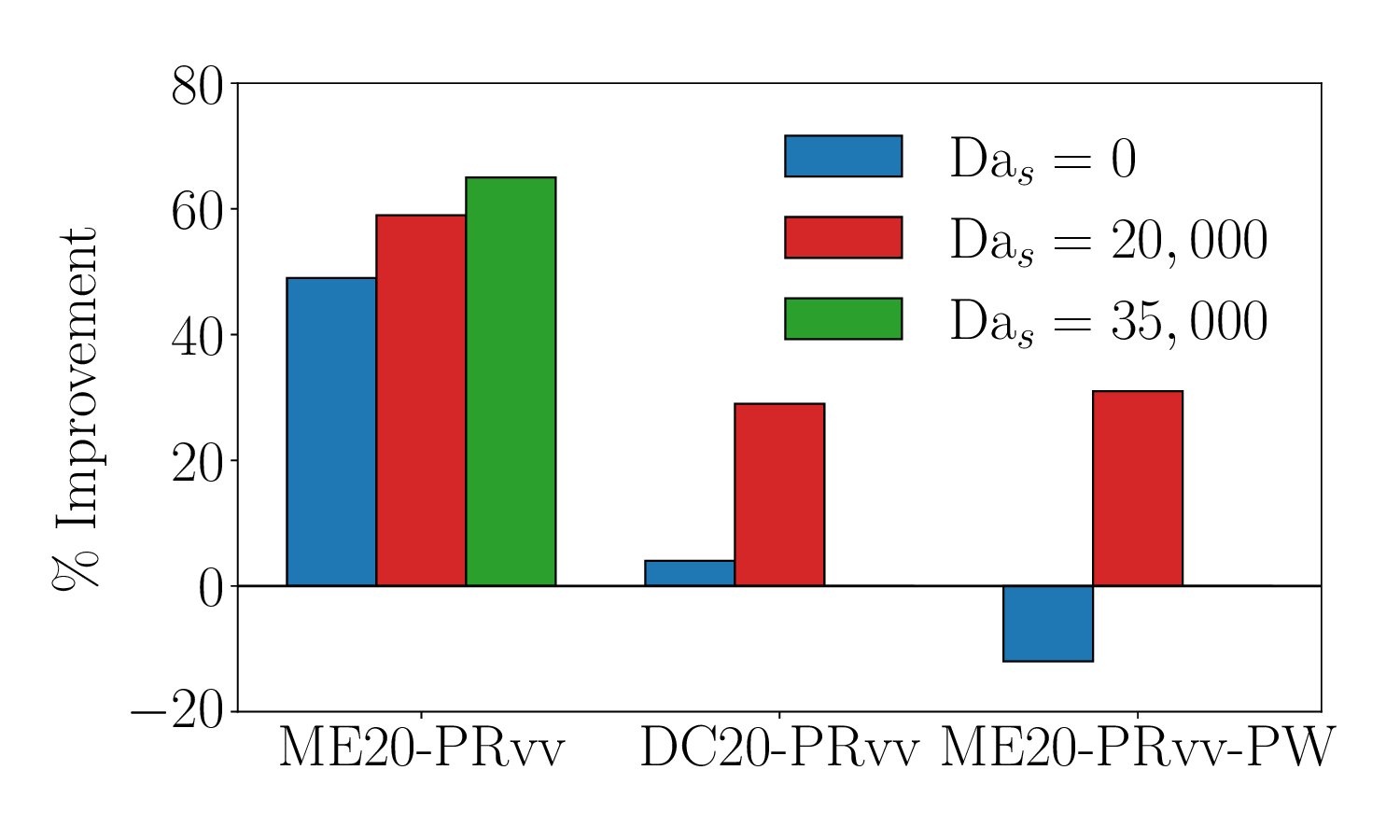}
   \caption{Relative percentage improvements of ME model variants compared to the Smagorinsky baseline. Red bars denote in-sample results ($\Da_s = 20,000$); all other bars show out-of-sample performance.}
\label{fig:model_comparison}
\end{figure}
\begin{figure}
    \centering
\includegraphics[width=0.82\linewidth]{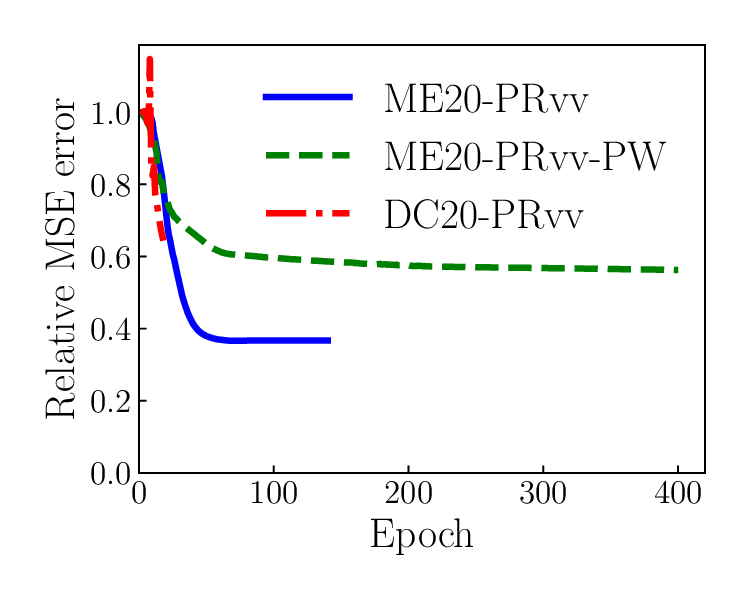}
    \caption{Training-loss curves comparing ME20-PRvv, DC20-PRvv, and ME20-PRvv-PW at $\Da_{s}=20{,}000$. Cumulative MSE error is normalized by the initial loss.}
    \label{fig:train_ME_PRvv}
    \vspace{-10pt}
\end{figure}
\subsection{Closure Model and Training Method Evaluation\label{subsec:Closure Model and Training Method Evaluation}}\addvspace{10pt}
The selection of an appropriate closure formulation is fundamental to balancing predictive accuracy with numerical robustness. Figure~\ref{fig:model_comparison} illustrates model improvement relative to the Smagorinsky baseline; the ME20-PRvv model delivers significantly better in-sample performance than the DC20-PRvv model when evaluated against the PRvv loss. This performance gap is primarily driven by the increased output dimensionality of the DC formulation, which requires the prediction of 15 separate terms. Compared to the more streamlined two-output ME closure, this structural complexity hinders both training convergence and the model's capacity for generalization across varying flow conditions. 

The robustness of the ME formulation is further confirmed through out-of-sample evaluations across varying Damk\"ohler number regimes. As shown in Fig.~\ref{fig:model_comparison}, the ME20-PRvv model maintains consistent performance gains, exceeding $40\%$ for the PRvv loss at both $\Da_{s}=0$ and $\Da_{s}=35{,}000$. Conversely, the DC20-PRvv formulation exhibits poor generalization; while it shows a modest improvement in the PRvv metric at $\text{Da}_{s}=0$, the underlying solutions for the primitive fields and other transport statistics are degraded relative to the baseline. Furthermore, the DC model becomes numerically unstable at $\text{Da}_{s}=35{,}000$, a failure that highlights the inherent sensitivity of high-dimensional direct-stress formulations in high Damk\"ohler number regimes. 

The impact of this high dimensionality on convergence is captured by the training loss histories. Shown in Fig.~\ref{fig:train_ME_PRvv}, the ME20-PRvv model converges steadily to a training loss reduction of approximately 64\%, reaching a stabilized relative MSE of approximately 0.36. In contrast, the DC20-PRvv model exhibits erratic training convergence, characterized by a significant initial error spike and only a limited 35.6\% reduction before numerical instability forces the premature termination of the optimization. This behavior primarily arises from the coupled PDE solve, in which certain learned closure-field combinations can drive the solution toward a numerically unstable state.

To further evaluate the training strategies for the ME model, both parallel-window (PW) and serial-window approaches were compared for the ME20-PRvv formulation. For the in-sample case at $\Da_{s}=20{,}000$, while both strategies succeed in reducing errors, the gain in $J_{PR_{vv}}$ for the serial model is nearly twice that of its parallel counterpart. Although parallel training improves the PRvv metric compared to the Smagorinsky baseline, it fails to extend these gains to other Reynolds stress components or scalar fluxes and in certain instances, degrades them relative to the baseline model. Out-of-sample results further substantiate this trend: at $\Da_{s}=0$, the ME20-PRvv-PW model underperforms the baseline across nearly all metrics, and at $\Da_{s}=35{,}000$, the simulation becomes entirely unstable.

As illustrated in Fig.~\ref{fig:train_ME_PRvv}, the training loss curves highlight the superiority of the serial approach, which achieves a loss reduction of $64\%$ compared to the $43\%$ achieved by the PW configuration. This disparity is rooted in the optimization logic: for the PW configuration, the target state is reinitialized at the start of each optimization window ($N_\mathrm{op}$), which means that errors only accumulate within single intervals. During deployment, however, the model must evolve continuously over the entire trajectory ($N_T$), for which the PW formulation is not explicitly trained. In addition, reinitializing each window from the filtered DNS may introduce a short adjustment as the LES evolves under its own closure and numerical discretization, adding noise to the parallel-window optimization. Consequently, despite exhibiting lower complexity during its localized optimization windows, the ME20-PRvv-PW model fails to preserve numerical stability during long-horizon simulations. These observations identify the serial-window strategy as the most robust training configuration; therefore, this approach is adopted for the subsequent analysis. 

In addition to the ME and DC closure formulations reported above, we also examined alternative scalar-closure formulations. First, the SGS diffusivity in~\eqref{eq:species_closure_eq} was predicted as an additional neural-network output rather than being determined from the turbulent viscosity through a turbulent Schmidt-type relation. This modification did not improve the \emph{a posteriori} accuracy or generalization of the model. Second, we attempted to model the unresolved reaction-source contribution as an additional output of the ME model. In this case, the network could not robustly separate the effects of unresolved turbulent transport and unresolved reaction in the species transport equation, or attribute these effects distinctly to the designated neural-network outputs. This led to less stable training without improved accuracy. In the present formulation, the unresolved reaction source is therefore not modeled explicitly, but is assumed to be indirectly represented through the learned species-transport closure. A more general treatment of the unresolved reaction source would likely require additional physical constraints to isolate the network prediction to the intended unresolved closure terms; this is the subject of ongoing research. We also do not substitute an established filtered chemical source closure, since the adjoint-based optimization would implicitly correct for any deficiencies, likely leading to similar improvements relative to the respective baseline LES.

\subsection{Effect of Loss Formulation\label{subsec:Effect of Loss Formulation}}\addvspace{10pt}

\begin{figure*}[t]
\vspace*{-0.25cm}
    \centering
    \includegraphics[width=0.9\linewidth]{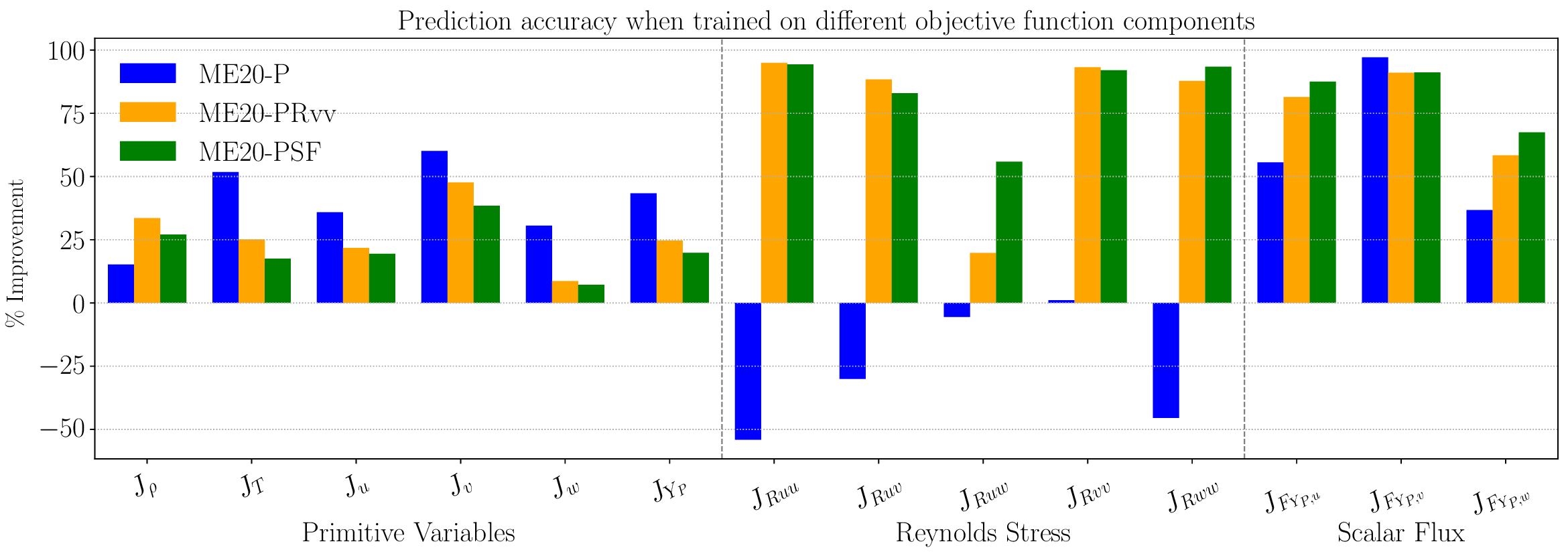}
   \caption{Component-wise validation losses for ME20-P, ME20-PRvv and ME20-PSF at $\Da_{s}=20{,}000$, expressed as percentage improvements relative to the Smagorinsky baseline.}
    \label{fig:bar_chart_for_diff_losses}
\end{figure*}
The impact of the different training objectives
defined in Section~\ref{subsec:objective_functions} for the in-sample cases is shown in Fig.~\ref{fig:bar_chart_for_diff_losses}, which reports component-wise validation losses as percentage changes relative to the Smagorinsky baseline. Results are grouped into primitive variables, resolved Reynolds stresses, and scalar fluxes evaluated at $\Da_s$ = 20,000, an intermediate case with significant flame–turbulence interactions. Three objective functions, ME20-P, ME20-PRvv and ME20-PSF, are examined to assess the effect of including second-order statistics in the training objective.

\begin{table*}[ht]
\centering
\caption{Percentage improvements of ME loss variants relative to the Smagorinsky baseline at scaling Damk\"ohler numbers $\Da_{s}=0,\,20{,}000,$ and $35{,}000$. Red highlights denote the in-sample cases at $\Da_{s}=20{,}000$.}
\label{tab:loss_variants_comparison}
\small
\begin{tabular}{c|ccc|ccc|ccc}
\toprule
\multirow{2}{*}{\textbf{Model}} &
\multicolumn{3}{c|}{$\Da_s = 0$} &
\multicolumn{3}{c|}{$\Da_s = 20{,}000$} &
\multicolumn{3}{c}{$\Da_s = 35{,}000$} \\
& $J_{P}$  & $J_{R_{vv}}$ & $J_{SF}$ 
& $J_{P}$  & $J_{R_{vv}}$ & $J_{SF}$ 
& $J_{P}$ & $J_{R_{vv}}$ & $J_{SF}$ \\
\midrule
ME20-P     & 1 & -81 & -60  & \cellcolor{red!20} 44 & 1 & -6   & 56 & 51 & 37 \\ 
ME20-PRvv  & -11 & 95 & 63  & \cellcolor{red!20}29 & \cellcolor{red!20}93 & 65   & 41 & 89 & 76 \\
ME20-PSF   & -12 & 93 & 65  & \cellcolor{red!20}23 & \cellcolor{red!20}91 & \cellcolor{red!20}79 & 34 & 87 & 70 \\
\bottomrule
\end{tabular}
\end{table*}
\indent The ME20-P model, trained solely on the primitive-variable loss, yields the largest mean-flow improvements. Velocity errors in the flame-normal direction ($J_v$) and temperature errors ($J_T$) both improve by over $50\,\%$.  
While this objective successfully improves scalar flux, it fails to recover second-order statistics; most resolved Reynolds stress components deteriorate relative to the baseline. These results highlight the limitation of a purely primitive-based objective, which accurately reproduces mean fields but fails to reconstruct the underlying Reynolds stress. 

Including second-order statistics in the objective largely alleviates these shortcomings. Both the ME20-PRvv and ME20-PSF models substantially improve the resolved Reynolds stress and scalar-flux accuracy while maintaining similar primitive variable accuracy to ME20-P. 
The ME20-PRvv model yields the most balanced improvements overall, with primitive errors such as $J_v$ and $J_T$ improving by approximately 50\% and 25\%, respectively, while simultaneously achieving error reductions exceeding 75\% for most second-order statistics. 

Table~\ref{tab:loss_variants_comparison} substantiates the selection of ME20-PRvv for robust, multi-regime applications and highlights its superior out-of-sample generalization. While the primitive-only ME20-P model fails to recover the resolved Reynolds stress in the unseen non-reacting limit ($\Da_s = 0$), resulting in an 81\% degradation, ME20-PRvv maintains $J_{R_{vv}}$ improvements exceeding 89\% and $J_{SF}$ gains above 63\% across all tested regimes. At the out-of-sample $\Da_s = 0$ condition, the marginal degradation in the primitive-variable loss relative to the baseline occurs because the standard Smagorinsky closure is already highly accurate for nonreacting shear flows. In the final comparison between the two second-order variants, ME20-PRvv is adopted as the preferred formulation. It provides more consistent improvements in the resolved Reynolds stress across the entire Damköhler number range and demonstrates superior scalar-flux recovery in the strongly reacting regime. 

\begin{figure*}[h]
\vspace*{-0.45cm}
    \centering
    \includegraphics[width=0.9\linewidth]{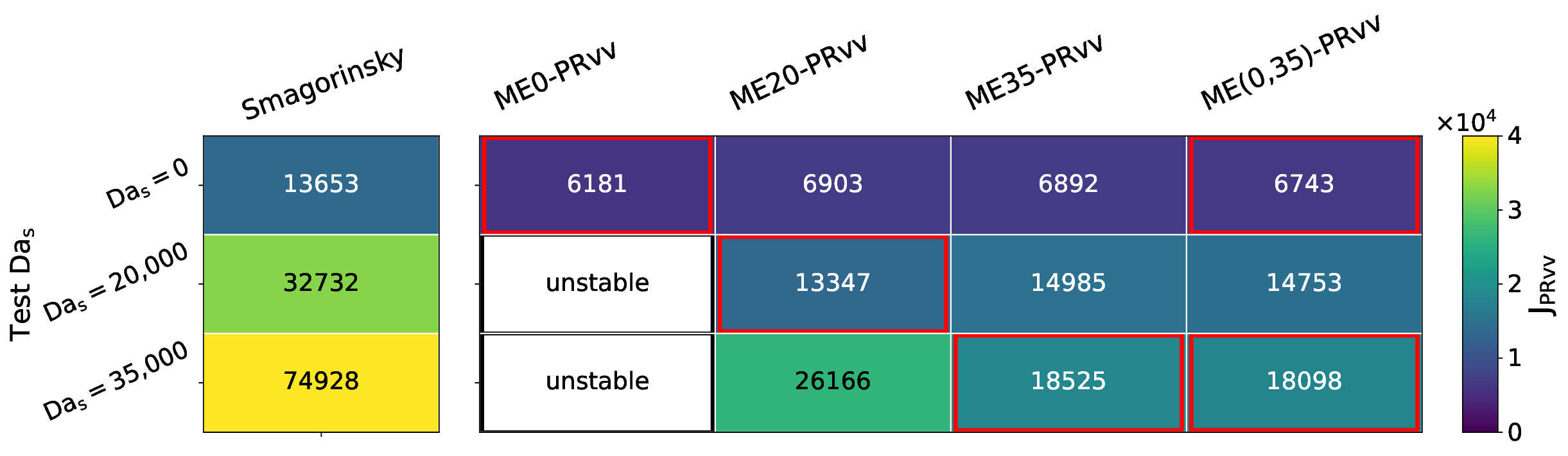}
    \caption{Performance of ME-PRvv models trained at different Damk\"ohler numbers and evaluated at $\Da_s=0,\,20{,}000,$ and $35{,}000$. Red boxes denote in-sample testing. }
    \label{fig:all_Da_comp}
\end{figure*}
\subsection{In- and Out-of-sample ME20-PRvv Predictions
\label{subsec:In-Sample and Out-of-sample Predictions with ME20-PRvv}}\addvspace{10pt}

\begin{figure}[!ht]
    \centering
    \includegraphics[width=0.9\linewidth]{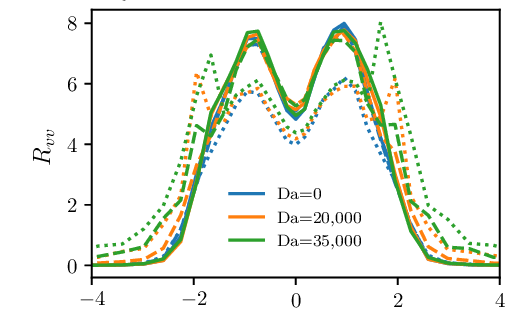}
    
    \vspace{1em} % Adds a small vertical gap between the images
    
    \includegraphics[width=0.9\linewidth]{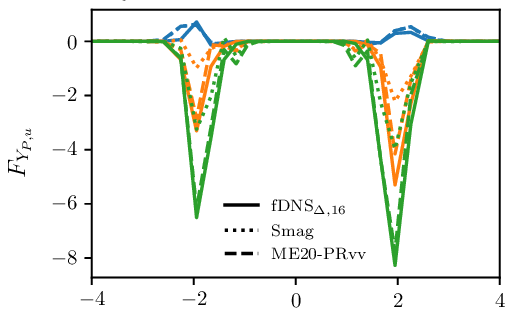}
    \caption{Comparison of Reynolds stress $R_{vv}$ and scalar flux $F_{Y_{P,u}}$ at $t=16$ for $\Delta_{16}$ at scaling Damk\"ohler numbers $\Da_s=0,\,20{,}000,$ and $35{,}000$ using the ME20-PRvv model.}
    \label{fig:out-sample_stress}
    \vspace{-10pt}
\end{figure}

To comprehensively evaluate the out-of-sample prediction capability of the ME-PRvv model, Fig.~\ref{fig:all_Da_comp} compares models trained at $\Da_s=0$, $20{,}000$, and $35{,}000$, along with the multi-case-trained model ME(0,35)-PRvv trained over $\Da_s=0$ and $35{,}000$, each tested for all three conditions.  
ME0-PRvv, trained exclusively on the nonreacting case, reduces $J_{PR_{vv}}$ by approximately $55\,\%$ at $\Da_s=0$ but becomes unstable under reacting conditions.  
Training at the intermediate regime yields broader generalization benefits: ME20-PRvv lowers $J_{PR_{vv}}$ by about $59\%$ at $\Da_s=20{,}000$ and $65\%$ at $\Da_s=35{,}000$.  
ME35-PRvv, trained in the strongly reacting regime, delivers the largest in-sample gains with $J_{PR_{vv}}$ improvements exceeding $75\%$, while exhibiting comparable performance to ME20-PRvv at $\Da_s=0$ and lower accuracy at $\Da_s=20{,}000$.  

Including multiple regimes in training further enhances robustness: ME(0,35)-PRvv achieves balanced performance, reducing $J_{PR_{vv}}$ by roughly $51\,\%$ at $\Da_s=0$, $55\,\%$ at $\Da_s=20{,}000$, and $76\,\%$ at $\Da_s=35{,}000$. 
The component-level generalization performance of the ME20-PRvv model is further evaluated under out-of-sample conditions.  Figure~\ref{fig:out-sample_stress} compares the normal stress $R_{vv}$ component and the streamwise scalar flux $F_{Y_{P,u}}$ component at $\Da_s=0$ (nonreacting jet) and $\Da_s=35{,}000$ (strongly burning flame) against DNS data.  
While the Smagorinsky baseline performs poorly in both limits, underpredicting $R_{vv}$ and  scalar-flux magnitudes, the ME20-PRvv maintains accuracy across the entire range of Damk\"ohler numbers, correctly capturing both the magnitude and the spread of the Reynolds stress while correctly predicting the magnitude and the direction of scalar transport even in unseen regimes.  
These results demonstrate the superior generalization capability of the ME20-PRvv model.    

\section{Computational Cost\label{sec:cost}}\addvspace{10pt}
We benchmark the cost of adjoint-based training using NVIDIA V100 GPUs (32 GB) and an AMD Ryzen Threadripper PRO 5945WX CPU (4.1 GHz, 12 threads). For an epoch of $N_T = 480$ steps, the training wall-clock time is 3720 s on CPU, 635 s on a single GPU, and 167 s on four GPUs using parallel windows. These represent GPU acceleration of approximately $6$ and $22$ times relative to the CPU cost. Total memory usage per LES step is approximately 4.2 GB, including all intermediate gradients and fluxes. A non-graph-breaking optimization approach would require storing this full state (for construction of the complete computational graph) over $N_\mathrm{op}=60$ steps, which would exceed each GPU's 32 GB memory limit requiring approximately 250 GB. 
Instead, the adjoint method stores the primitive variables and breaks the computational graph at each time step using Eq.~\ref{eq:grad_L}, ensuring constant memory requirements regardless of $N_{\text{op}}$. This avoids the linear memory growth associated with non-graph-breaking PDE-constrained optimization approaches, with the tradeoff that the adjoint solve accounts for approximately 70\% of the total cost within the optimization loop. Overall, this method ensures LES consistency while enabling scalable training at tractable memory cost.

We evaluate the inference cost of the trained closure separately from the adjoint-based training cost. During inference, only the forward LES solve and neural network closure evaluation are required; no adjoint integration or backpropagation is performed. For the $N_T=480$-step trajectory, deep learning-integrated LES inference requires 109.56~s on an NVIDIA V100 GPU (peak memory usage of 2183.36 MB) or 642~s on a single CPU core. The corresponding Smagorinsky LES requires 82.54~s on the GPU (peak memory usage of 732 MB) or 100.52 s on a single CPU core. The neural network's evaluation is very efficient on the GPU,  leading to its modest overhead versus the Smagorinsky baseline for GPU execution.

\section{Conclusion\label{chap:conclusion}}\addvspace{10pt}

This work extends the Deep Learning PDE Model (DPM) framework to LES of turbulent premixed jet flames. By embedding a neural network closure directly within the governing equations, the augmented eddy-viscosity formulation achieves stable convergence and accurate subgrid transport predictions across varying Damköhler numbers.

The augmented eddy-viscosity architecture demonstrates superior stability and generalization compared to the high-dimensional direct-closure approach. While the direct-closure model offers theoretically greater flexibility, its 15-term output dimensionality leads to numerical instability in reacting regimes. Furthermore, the serial-window training strategy is identified as the most robust approach for reacting flows due to its correction of long-time error growth. This method significantly outperforms parallel-window configurations used in previous non-reacting works.

Supervised training to match second-order filtered-DNS  statistics is crucial for capturing turbulence dynamics, with models trained only for primitive variables failing to recover resolved fluctuations. While incorporating a comprehensive set of second-order metrics in the objective function provides significant supervision, the formulation focusing on flame-normal stress components emerges as the preferred model. This choice offers the most consistent balance between mean-field accuracy and resolved turbulence statistics across all tested regimes. Ultimately, this framework demonstrates that PDE-consistent deep learning closures provide a robust foundation for multi-regime turbulent combustion modeling. 

Future research will extend the DPM framework to multispecies thermochemistry, reaction-source-term closures, and transport modeling for complex fuels. Its applicability will be further assessed by generalizing the model across complex geometries and a broader range of turbulent combustion regimes, with systematic benchmarking against advanced subgrid-scale closures and established turbulence--chemistry interaction models, as well as detailed evaluation of combustion-related metrics.
\acknowledgement{CRediT authorship contribution statement} \addvspace{10pt}

{\bf Priyesh Kakka}: Conceptualization, Methodology, Formal analysis, Investigation, Visualization, Writing - Original Draft,  Writing - Review \& Editing. {\bf Jonathan F. MacArt}: Conceptualization, Methodology, Investigation, Writing - Review \& Editing, Supervision, Funding acquisition.

\acknowledgement{Declaration of competing interest} \addvspace{10pt}

The authors declare no competing financial interests or personal relationships that could have appeared to influence the work reported in this paper.

\acknowledgement{Acknowledgments} \addvspace{10pt}

The authors gratefully acknowledge support from the U.S.\ NSF under Award CBET-2236904.

% -------------------------------------------------------------------- %
% -------------------------------------------------------------------- %
% -------------------------------------------------------------------- %
\footnotesize
\baselineskip 9pt

% -------------------------------------------------------------------- %
% -------------------------------------------------------------------- %
% -------------------------------------------------------------------- %

\thispagestyle{empty}
\bibliographystyle{latex}
\bibliography{LaTeX}

% -------------------------------------------------------------------- %
% -------------------------------------------------------------------- %
% -------------------------------------------------------------------- %

\small
\baselineskip 10pt
% -------------------------------------------------------------------- %
% -------------------------------------------------------------------- %
% -------------------------------------------------------------------- %

\end{document}